\documentclass[twocolumn,showpacs,preprintnumbers,amsmath,amssymb,floatfix]{revtex4}
\usepackage{graphicx}
\usepackage{afterpage}
\usepackage{amsmath}
\usepackage{dcolumn}
\usepackage{bm}
\usepackage{nicefrac}

\graphicspath{{Biezums_img/}{Jauda_img/}{Trans_img/}{Bildes/}{./}{Temp_img/}}

\begin{document}

\preprint{APS/123-QED}

\title{Relaxation mechanisms affecting magneto-optical resonances in an extremely thin cell: experiment and theory for the cesium D$_1$ line}

\author{M.~Auzinsh$^1$}
\email{Marcis.Auzins@lu.lv}
\author{A.~Berzins$^1$}%
\author{R.~Ferber$^1$}%
\author{F.~Gahbauer$^1$}%
\author{U.~Kalnins$^1$}
\author{L.~Kalvans$^1$}%
\author{R.~Rundans$^1$}%
\author{D.~Sarkisyan$^2$}%
\affiliation{ $^1$Laser Centre, The University of Latvia, 19 Rainis
Boulevard, LV-1586 Riga, Latvia\\
$^2$Institute for Physical Research, NAS of Armenia, Ashtarak-0203, Armenia}%

\date{\today}

\begin{abstract}
We have measured magneto-optical signals obtained by exciting the $D_1$ line of cesium atoms confined to an extremely thin cell (ETC), 
whose walls are separated by less than one micrometer, and developed an improved theoretical model to describe these signals with 
experimental precision. 
The theoretical model was based on the optical Bloch equations and included all neighboring hyperfine transitions, the mixing of 
the magnetic sublevels in an external magnetic field, and the Doppler effect, as in previous studies. However, in order to 
model the extreme conditions in the ETC more realistically, the model was extended to include a unified treatment of 
transit relaxation and wall collisions with relaxation rates that were obtained directly from the 
thermal velocities of the atoms and the length scales involved. Furthermore, the interaction of the atoms with different regions 
of the laser beam were modeled separately to account for the varying laser beam intensity over the beam profile as well as 
saturation effects that become important near the center of the beam at the relatively high laser intensities used 
during the experiments in order to obtain measurable signals. The model described the experimentally measured signals 
for laser intensities for magnetic fields up to 55~G and laser intensities up to 1~W/cm$^2$ with excellent agreement.  
\end{abstract}
\maketitle

\section{\label{Intro:level1}Introduction}
Modern fabrication techniques and technological requirements have led to a growing demand for
smaller and smaller atomic vapor cells~\cite{Mhaskar:2012}. These "labs-on-a-chip" find applications in distributed atomic 
clocks that are required for precise global positioning, among other applications. Furthermore, 
extremely thin cells (ETCs)~\cite{Sarkisyan:2001}, one of whose dimensions can be as thin as several tens of nanometers, 
have found applications in fundamental research because they allow sub-Doppler spectroscopy without the need for 
expensive or complicated setups such as atomic beams or magneto-optical traps~\cite{Sargsyan:2008}. 

Such small vapor cells 
present particular challenges to theoreticians who wish to model the experimental signals. For one, relaxation rates are 
extremely high as a result of collisions with walls. Furthermore, since the volume of vapor interacting with the laser radiation 
in the thin cell is very small and thus contains fewer atoms than usual, experiments are usually performed 
at high temperature and/or laser intensity. In the former case, reabsorption may begin to play a role. In the 
latter case, the atom-laser interaction becomes saturated in the central regions of the beam, which means that the 
transit relaxation cannot be described any longer by an exponential with a single decay constant~\cite{Auzinsh:1983}.  

In this study, we aim to show that a precise description of magneto-optical signals in an ETC can be obtained with a 
theoretical model that is based on the optical Bloch equations extended to include a more detailed treatment of relaxation 
processes and the saturation of the atom-laser interaction in the high-intensity areas of the beam. 

We focused our study on the magneto-optical resonances, which are closely related to the ground-state 
Hanle effect, first observed in cadmium~\cite{Lehmann:1964}. These resonances can be observed, when at 
zero magnetic field a dark superposition state that does not absorb laser radiation forms among various 
degenerate ground-state magnetic sublevels~\cite{Kazantsev:1984, Renzoni:2001a,Alnis:2001}. This coherent dark state 
is destroyed when a magnetic field is applied, leading to an increase of absorption and laser induced fluorescence (LIF). 
So-called dark resonances had been described earlier in alkali atoms~\cite{Schmieder:1970, Alzetta:1976}. 
The Bloch equations were first applied to modeling dark resonances in a sodium beam~\cite{Picque:1978}. 
In the linear regime at low powers analytical descriptions of such resonances are possible~\cite{Breschi:2012}. 
However, as the absorption of laser radiation becomes nonlinear, numerical models are normally used, and over time these 
models have been extended to include the coherent properties of the exciting laser radiation, 
the simultaneous interaction of all hyperfine sublevels in the ground and excited states with the laser radiation, 
magnetic-field-induced mixing of magnetic sublevels, and the Doppler effect~\cite{Auzinsh:2008}. 
These magneto-optical resonances provide a good test case for theoretical models as they include a wide 
range of coherent and incoherent processes~\cite{Auzinsh:2013}. 

Optical signals obtained with ETCs, where alkali vapor is confined between two YAG crystals separated by 
a few hundred nanometers, present interesting properties related to the small dimension of the 
cell in the direction along a laser beam that is perpendicular to the ETC walls. 
Early efforts to model these signals focused on the Dicke-type 
narrowing observed in transmission signals when the wall separation was equal to an odd number of half 
wavelengths~\cite{DutierJOSA:2003}. However, for fluorescence signals, this effect is not expected to be 
significant~\cite{Maurin:2005}. The effect of energy shifts induced by van der Waals interactions with the walls was 
considered, and for the cesium $D_1$ line could lead to shifts of the excitation maximum by up to 200 MHz for cell 
thicknesses around 50 nm~\cite{Hamdi:2005}. Some of these early studies also applied the optical Bloch equations 
to a simplified two-level representation with optical pumping losses~\cite{Sargsyan:2008b, DutierJOSA:2003}, 
but only in the linear excitation regime for intensities up to several milliwatts per square centimeter and 
Rabi frequencies less than 10 MHz. More recently, the cesium $D_2$ line has been studied experimentally and 
theoretically at somewhat higher powers up to 40 mW/cm$^2$ using a model based on the optical Bloch 
equations~\cite{Andreeva:2007, Cartaleva:2013}, but these models still were based on an open two-level system 
rather than including all degenerate sublevels. Furthermore, theoretical results were not compared directly 
to experimental curves, with the exception of some aggregate parameters, such as peak amplitude. 

More sophisticated models were applied to describe experimentally measured magneto-optical signals obtained 
for the cesium $D_2$~\cite{Andreeva:2007b} and rubidium $D_1$~\cite{Auzinsh:2010} lines for atoms in ETCs, 
including all degenerate hyperfine levels of the excited- and ground-state manifolds. However, the proper treatment 
of collisions remained difficult and somewhat contradictory. The study of cesium atoms 
postulated a model for elastic collisions that redistributed the excited state populations 
with weights that depended on the transition strength. In contrast, the study of rubidium atoms 
excluded these elastic collision terms and focused only on relaxation that was based on wall collisions. However, 
the best agreement between experiment and theory could only be obtained when different collisional relaxation rates were 
assumed for the ground and excited states. 

We now applied a theoretical model to describe fluorescence signals of magneto-optical resonances that built 
on the previous efforts and added a more comprehensive approach towards treating properly saturation effects that are important 
for intense laser radiation whose power density varies over the beam profile. 
As in previous studies, the Bloch equations were applied, and all degenerate sublevels were included. 
The mixing of magnetic sublevels in an external magnetic field was also included, because these become important 
for magnetic fields of several tens of Gauss or more. The equations were solved for various velocity groups 
and averaged over the residual Doppler profile. The equations contained relaxation terms for wall collisions 
and transit relaxation that were derived from the thermal velocities and the dimensions of the cell and the 
laser beam. 

The new feature included in the model this time to account for saturation effects and the varying 
laser intensity over the beam profile consisted in dividing the laser beam into concentric rings, 
and solving the Bloch equations separately for each ring, summing up the  
results with the appropriate weights. The particle exchange due to thermal motion between these concentric regions was 
included in the theoretical description. This final addition to the model was crucial to improving substantially the 
agreement between experimentally measured signals and the theoretical curves for magnetic field values up to at least 55 G, 
and it is a step towards resolving a outstanding problem in the modeling of magneto-optical signals~\cite{Auzinsh:2008}. 

In this article we present new measurements of magneto-optical resonances recorded for cesium atoms confined to 
an ETC. Resonances have been recorded for all transitions of the $D_1$ line of cesium. 
In contrast to the complexity of the Rb $D_1$ or $D_2$ lines, the Cs $D_1$ line is much simpler. 
In this system, the increased hyperfine splitting and the presence of only one isotope make possible 
the investigation of individual hyperfine transitions. 
The ground state hyperfine splitting is over 9~GHz and the excited state hyperfine splitting 
is 1.17~GHz (see Fig.~\ref{fig:levels}, markedly exceeding the Doppler width.
The temperature dependence of the resonance 
shapes has been studied for temperatures up to 165$^{\circ}$C. Furthermore, the dependence of the resonance shapes on laser power and 
ETC wall separation have been studied. The extended theoretical model has been applied to describe the experimental signals with excellent 
results.

\section{\label{Experiment:level1}Experiment}

The ETC~\cite{Sarkisyan:2001} consists of two YAG crystal windows that are glued together in such 
a way that the distance between them varies from 50~nm to 1.2~$\mu$m.
It was produced at the Institute of Physical Research in Ashtarak, Armenia.
The cesium is stored in a separate reservoir that is connected to the bottom part of the ETC.
An additional thermocouple is attached to the cesium container for temperature measurements. 
For a more detailed description see~\cite{Sarkisyan:2001}. 

\begin{figure}
	\includegraphics[width=0.45\textwidth]{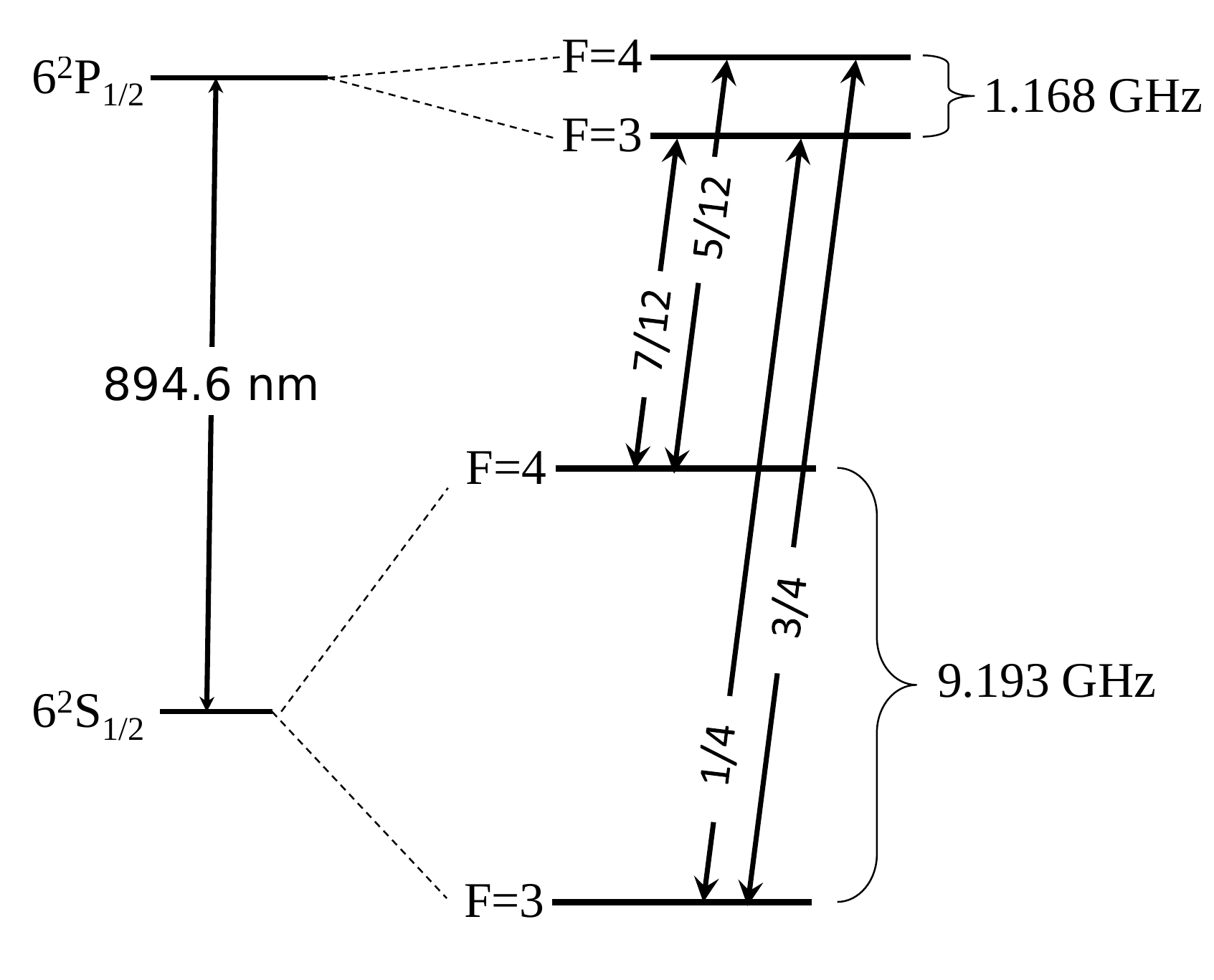}
    \caption{\label{fig:levels} Hyperfine level structure and transitions of the  $D_1$ line of cesium (not to scale). 
The fractions on the arrows indicate the relative transition strengths.}
\end{figure}

\begin{figure}
	\includegraphics[width=0.45\textwidth]{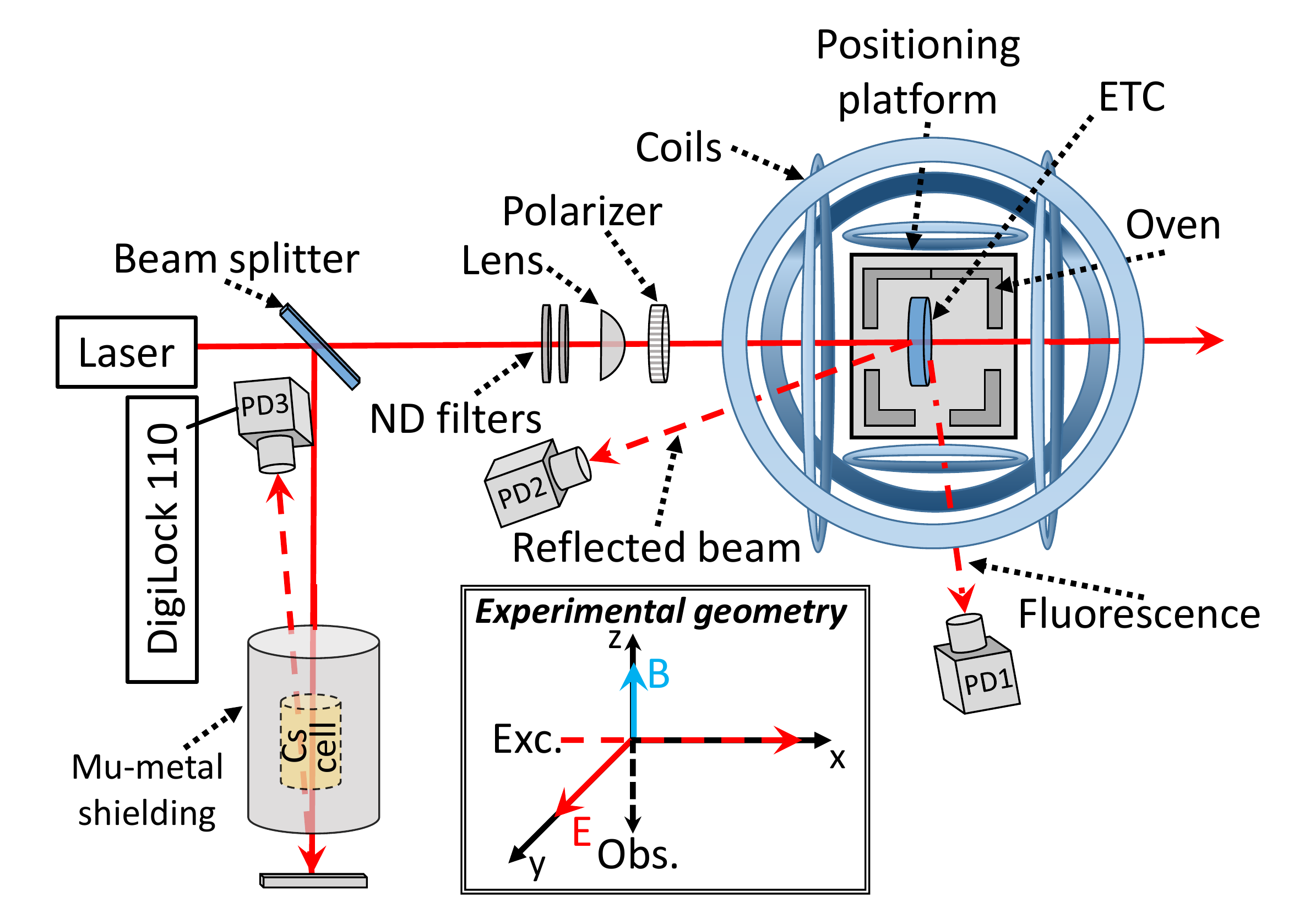}
	\caption{\label{fig:exp_setup} Schematic representation (top view) of the experimental setup.}
\end{figure}

In this experiment we performed magnetic field scans over the zero-field resonance up to 55 G and detected 
the fluorescence signal from the ETC. We measured zero field resonances for all four possible transitions of the 
Cs $D_1$ line (see Fig.~\ref{fig:levels}).  
The experimental setup is shown in Fig.~\ref{fig:exp_setup}.

 To control the temperature within the ETC and increase the vapour pressure, the cell was positioned 
inside a small oven. This oven was heated by a current that flowed through a bifilar coil in opposite directions to allow heating 
without producing stray magnetic fields. The oven consisted of two separate heating elements, which 
made it possible to maintain the cell itself and the cesium container at different temperatures. The cell's temperature was maintained at
about 195$^\circ$C, while the temperature of the reservoir was varied between 70$^\circ$C and 165$^\circ$C. 
Both the cell and the oven were placed on top of a Thorlabs NanoMax table, which allowed precise 3D positioning through three 
computer controlled motors, each offering 4 mm of scan range. This setup was positioned at the center of three pairs of Helmholtz coils. 
Two coils were used to compensate the ambient magnetic field, and the third coil was used to scan the magnetic field up to 55~G 
by applying a current from a Kepco BOP-50-8-M bipolar power supply, which was controlled by a TTI TG5011 function generator.	
	
The atoms were excited by a diode laser (DL100-DFB) from Toptica, A.G., based on a distributed feedback laser diode 
appropriate for exciting the $D_1$ line of cesium. The laser beam was split into two beams by a beam splitter. The weaker laser beam 
was directed to the cesium reference cell that was placed inside a three-layer mu-metal shield to eliminate the magnetic fields. 
The saturation absorption signal from the reference cell (measured by photodiode PD3) was used in conjunction with the Toptica 
“feedback controlyzer DigiLock 110” for laser frequency stabilization. 

The laser beam power was adjusted using a series of 
neutral density filters, and the laser beam was focused onto the ETC with a lens whose focal length was 50 cm. 
The laser beam diameter was measured to be approximately 200~$\mu$m.  
Linear polarization of the laser beam (electric field vector $\bm E$ in Fig.~\ref{fig:exp_setup}) 
was achieved with a Glan-Thompson polarizer (Thorlabs GTH10M). 
Because the glue reduces the amount of light that can pass through the side of the ETC, 
the fluorescence signal from ETC was collected in a direction that made a small angle with the ETC cell walls 
and was focused using a series of lenses onto photodiode PD1 (Thorlabs FDS100). 
The lens system was constructed in such a way that it collected the fluorescence signal from only a small region, 
ignoring most of the scattered light from the cell walls. 

The intensity of the beam reflected back from the second wall of the ETC was measured using photodiode PD2 (Thorlabs PDA36A-EC). 
The measured intensity was used to detect the interference spectrum, which, in turn, was used to determine the cell thickness. 
The voltage signals from the photodiodes were recorded using digital oscilloscopes (Yokogawa DL-6154 and Tektronix TDS 2004B).

\section{\label{Theory:level1}Theory}

The basic approach of our theoretical description of magneto-optical signals has previously been described 
in~\cite{Auzinsh:2008} for ordinary vapor cells and expanded for the case of ETCs in~\cite{Auzinsh:2010}. 
In this section we describe how we extend the theoretical model to include a better description of the effects of transit relaxation and atom-wall collisions.

\begin{figure}
	\includegraphics[width=0.4\textwidth]{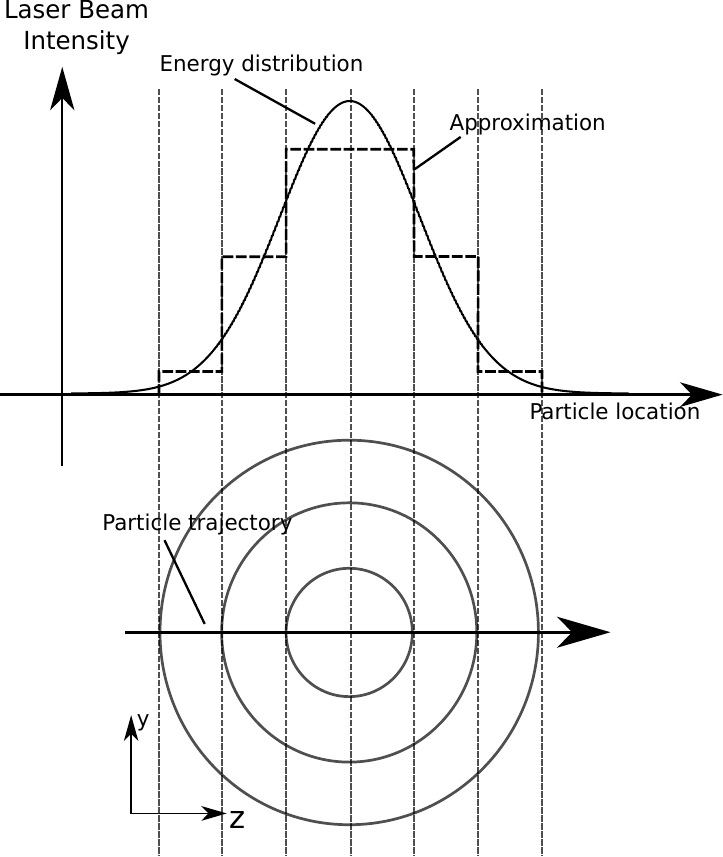}
\caption{\label{fig:dal1} Laser beam profile split into a number of concentric regions.}
\end{figure}

To derive rate equations that took into account the energy distribution within the laser beam profile, we assume that the particles 
move thermally, and their trajectories take them from outside the laser beam interaction region through the center of the laser beam. 
This assumption is somewhat simplistic, but can be easily extended by an averaging over multiple possible paths. For the purposes of this 
derivation we will assume that the laser beam is directed along the $x$-axis (Fig.~\ref{fig:exp_setup} inset), 
which means that particle movement can be split into two cases: 
movement in the $x$-direction, which accounts for Doppler broadening, and movement in the $yz$-plane, 
which accounts for transit relaxation. Even though particles in $yz$-plane move in random directions, 
the axial symmetry of the laser beam allows us to describe them with single trajectory for which the particle enters the laser beam 
from one side, moves along the $z$ axis and exits through the other side (Fig.~\ref{fig:dal1}).

To include energy distribution of the laser beam in our model we split the $z$-axis into multiple regions, 
giving each region its characteristic laser beam intensity (Fig. \ref{fig:dal1}). This means that rate equations 
have to be written and solved for each region separately, giving a separate density matrix for each one.

To derive the rate equations for a single region we start with the rate equations for the density matrix $\rho$ 
under the action of a Hamiltonian $\hat{H}$ and relaxation operator $\hat{R}$ in the following form: 
\begin{equation}
i \hbar \frac{\partial \rho}{\partial t} = \left[\hat{H},\rho \right]+ i \hbar \hat{R}\rho.
\end{equation}
To include particles leaving and entering the interaction region we add the following terms to the right-hand side of the equation:
\begin{equation}
-i \hbar\hat{\gamma_t} \rho+i \hbar\hat{\gamma_t} \rho',
\end{equation}
where $\rho'$ is the density matrix of particles entering the region and $\hat{\gamma_t}$ is the transit relaxation operator, 
which can be written as a diagonal relaxation matrix with elements 
${\gamma_t}_{ij}=v_{yz}/s_n \delta_{ij}$,
where $s_n$ is the region size and $v_{yz}$ is the characteristic particle speed. 
As we have assumed that all particles move in the same direction and transit regions sequentially, we can assume that for 
the $n$-th region $\rho'=\rho^{(n-1)}$. By adding terms that describe collisional relaxation with rate $\gamma_{c}$, 
we arrive at rate equations for the density matrix $\rho^{(n)}$ of the $n$-th region as:
\begin{align}
i~\hbar \frac{\partial\rho^{(n)}}{\partial t} &= \left[ \hat{H},\rho^{(n)}\right]+i ~\hbar \hat{R} \rho^{(n)} -i~\hbar\hat{\gamma_t}^{(n)}\rho^{(n)} \notag \\ 
& +i~\hbar\hat{\gamma_t}^{(n)}\rho^{(n-1)}-i~\hbar\hat{\gamma_c}\rho^{(n+i)}~\hbar\hat{\gamma_c}\rho^0,
\end{align}
where $\rho^{(0)}$ is the density matrix for atoms that do not interact with the laser radiation: 
their population is equally distributed among the ground-state sublevels. 
For atoms in the ETC there are two sources of collisional relaxation: atom-atom collisions and atom-wall collisions.
To include both effects in the theoretical model, we write the collision relaxation operator as a diagonal matrix with elements:
\begin{equation}
\label{collisions}
	\gamma_c= \frac{\bar{v}}{\bar{l}} + \frac{v_{x}}{L/2},
\end{equation}
where $\bar{v}$ is the mean atomic velocity, $\bar{l}$ is the mean free path, $v_{x}$ is the atomic velocity in the 
direction perpendicular to the cell walls, $L$ is the cell thickness.	
The mean free path $\overline{l}=\nicefrac{\overline{v}}{\Gamma_{SE}}$, where $\overline{v}$ 
is the mean thermal velocity of the cesium atoms, and 
the rate of spin-exchange $\Gamma_{SE}=\sigma_{SE}\overline{v}_{rel}n_a$, 
where $\sigma_{SE}=2.2\times10^{-14}$~cm$^2$ is the spin-exchange cross-section~\cite{Ernst:1968}, 
$\overline{v}_{rel}$=$(\nicefrac{8k_BT}{\pi \mu_{Cs}})^{1/2}$ is the average relative velocity of the cesium atoms 
($\mu_{Cs}$ is the reduced mass of the system of two cesium atoms), 
and $n_a$ is the density of atoms in the cell. At the temperatures studied here, the rate of atom-atom collisions is small 
compared to that of atom-wall collisions. 

By following the derivation of rate equations for Zeeman coherences described earlier~\cite{Blushs:2004} and neglecting transit relaxation of 
optical coherences we arrive at the following differential equations:
\begin{align}
\frac{\partial \rho_{g_i,g_j}^{(n)}}{\partial t} =& \sum_{e_k,e_m}\left(\Xi_{g_ie_m}^{(n)} + (\Xi_{e_kg_j}^{(n)})^*\right) d_{g_ie_k}^*d_{e_mg_j}\rho_{e_ke_m}^{(n)} \notag \\
& - \sum_{e_k,g_m}(\Xi_{e_kg_j}^{(n)})^*d_{g_ie_k}^*d_{e_kg_m}\rho_{g_mg_j}^{(n)} \notag \\ 
& - \sum_{e_k,g_m}\Xi_{g_ie_k}^{(n)} d_{g_me_k}^*d_{e_kg_j}\rho_{g_ig_m}^{(n)}  \\
& - i\omega_{g_ig_j}\rho_{g_ig_j}^{(n)}+\sum_{e_ke_l}\Gamma_{g_ig_j}^{e_ke_l}\rho_{e_ke_l}^{(n)}-\gamma_{t}\rho_{g_ig_j}^{(n)} \notag \\ 
& + \gamma_{t}\rho_{g_ig_j}^{(n-1)}-\gamma_{c}\rho_{g_ig_j}^{(n)}+\gamma_c\rho_{g_ig_j}^{(0)}\notag\\
\frac{\partial \rho_{e_i,e_j}^{(n)}}{\partial t} =& \sum_{g_k,g_m}\left((\Xi_{e_ig_m}^{(n)})^* + \Xi_{g_ke_j}^{(n)}\right) d_{e_ig_k}^*d_{g_me_j}\rho_{g_kg_m}^{(n)} \notag\\
& - \sum_{g_k,e_m}\Xi_{g_ke_j}^{(n)}d_{e_ig_k}d_{g_ke_m}^*\rho_{e_me_j}^{(n)} \notag \\ 
& - \sum_{g_k,e_m}(\Xi_{e_ig_k}^{(n)})^*d_{e_mg_k}d_{g_ke_j}^*\rho_{e_ie_m}^{(n)} \\
& - i\omega_{e_ie_j}\rho_{e_ie_j}^{(n)}-\Gamma\rho_{e_ie_j}^{(n)}-\gamma_{t}\rho_{e_ie_j}^{(n)}  \notag \\
& +\gamma_{t}\rho_{e_ie_j}^{(n-1)}-\gamma_{c}\rho_{e_ie_j}^{(n)} \notag,
\end{align}
where
\begin{align}
\Xi_{g_ie_j}= \frac{|\boldsymbol{\varepsilon}^{(n)}|^2}{\frac{\Gamma+\Delta\omega}{2}+i \left(\bar{\omega}-k v_x+\omega_{g_ie_j}\right)}
\end{align} 
characterizes the laser beam interaction with $|\boldsymbol{\varepsilon}^{(n)}|^2$, which is proportional to the laser beam intensity of the 
$n$-th region, $\Gamma$ is the spontaneous relaxation rate, $\Delta\omega$ is the spectral width of the laser beam, $\bar{\omega}$ is
the laser frequency, $kv_x$ is the Doppler shift for light with wave vector $k$ propagating in the $x$ direction,
$\omega_{g_ie_j}$ is the energy difference between the ground-state magnetic sublevel $g_i$ and the excited-state magnetic 
sublevel $e_j$,
and $d_{g_ie_j}$ and $d_{e_jg_i}$ are the dipole transition matrix elements that couple the 
excited- and ground-state magnetic sublevels. 
In order to relate the dipole strengths to laser intensity more conveniently, 
we can introduce the reduced Rabi frequency as $\Omega_R=\boldsymbol{\varepsilon}_r'/\hbar$, 
where $\boldsymbol{\varepsilon}_r'$ is the 
electric field intensity at the center of the laser beam and $d$ is the reduced matrix element of the dipole transition obtained 
from the matrix elements of $d_{e_jg_i}$ when the Wigner-Eckart theorem is applied~\cite{Blushs:2004}. 
Because the equations include a term for particle speed, the subsequent averaging of the fluorescence signal over the Maxwell 
distribution in the direction of the laser beam is needed to include Doppler shifts. 
An extra averaging over the $v_{yz}$ plane can be applied for the averaged transit relaxation, 
but the relative gains in precision are negligible at the laser powers of interest to us compared to the additional time required for the 
calculations. 
As the relaxation process dynamics are included in the transit relaxation terms, the calculations can be done for the steady state by 
solving a system of linear equations for density matrix $\rho$ that in turn can used to calculate fluorescence intensity 
$I\left(\boldsymbol{\varepsilon}_f\right)$ where $\boldsymbol{\varepsilon}_f$ is fluorescence polarization.  
	
In the end the fluorescence signal for polarization $\boldsymbol{\varepsilon}_f$ can be expressed as:
\begin{align}
	I(\boldsymbol{\varepsilon_f}) = \sum_{n} \sum_{v_x} f(v_x)\Delta v_x \frac{A_{n}}{A} I_{n}(v_x,\boldsymbol{\varepsilon}_f),
\end{align}
where $n$ runs over regions in the direction of the particle trajectory,  
$v_x$ is the particle speed in the direction of the laser beam, 
$f(v_x)\Delta v_x$ is the probability of finding atoms with velocity $v_x\pm \nicefrac{\Delta v_x}{2}$,  $\nicefrac{A_{n}}{A}$ is the 
relative area of region $n$ and $I_n(v_x,\boldsymbol{\varepsilon}_f)$ is the fluorescence intensity of the 
$n$-th region with atomic velocity $v_x$ in laser beam direction.    

\section{\label{Results:level1}Results and Discussion}

\begin{figure}[htpb]
	\includegraphics[width=0.4\textwidth]{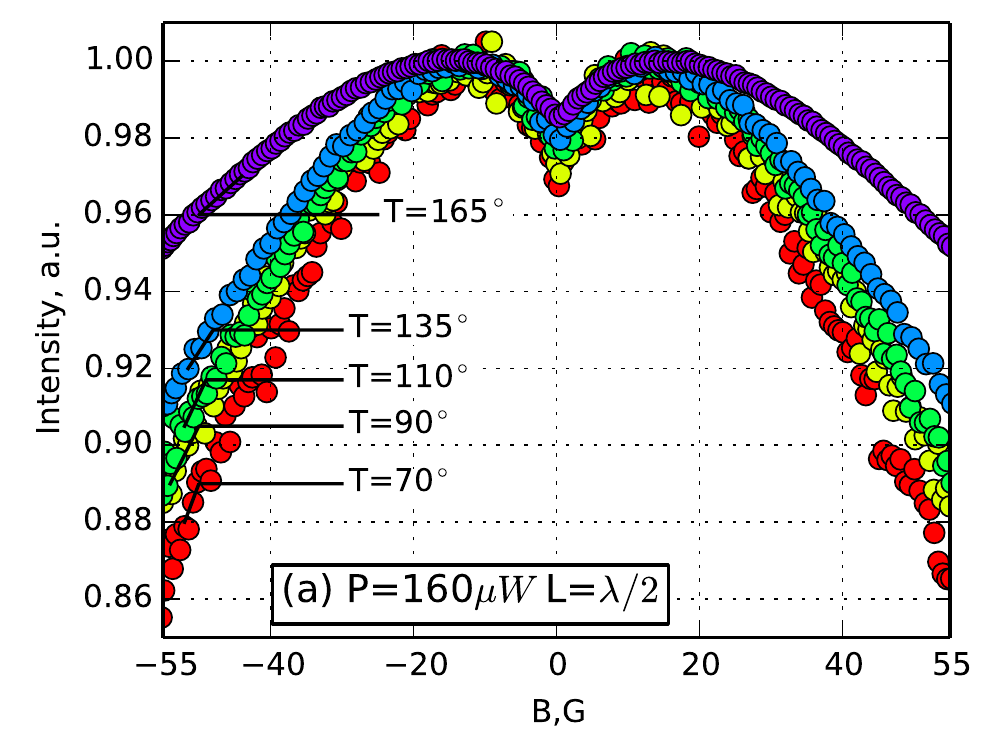}
	\caption{Signal dependence on temperature for $F_g=4\rightarrow F_e=4$ transition at cell wall separation of approximately 450~nm.}
	\label{fig:temp_dep}
\end{figure}

To analyze effects of the ETC on magneto-optical signals we began by measuring the temperature dependence of the 
Cs $D_1$ $F_g=4\rightarrow F_e=4$ transition for a wall separation of $L=\lambda/2$ and a constant laser power $P$ of 160~$\mu$W, 
which corresponds to an average power density of 500~mW/cm$^2$ for a laser beam diameter of 0.2~mm. 
The experimental results are shown in Fig. \ref{fig:temp_dep}. 
It can be seen that the flourescence signal recorded at 165$^{\circ}$C significantly deviates from signals obtained at
lower temperatures. 
The effect can be explained by the rapid vapour pressure rise in the cell, which 
eventually leads to reabsorption effects becoming important. Indeed, using the resonant 
cross-section of the $D_1$ line for $\pi$-polarized radiation and the density of cesium at 165$^{\circ}$C~\cite{Steck:cesium} 
one obtains an optical pathlength of 1.5~$\mu$m, 
which becomes comparable to the cell thickness and corresponds to an optical depth (OD) of 0.6.  
At similar optical depths (OD~$\sim 0.66$), reabsorption effects have been observed 
to start to play a role in an ordinary cell~\cite{Auzinsh:2012,Matsko:2002}. 
To study ETC effects without the added complication of reabsorption, we chose to conduct experiments at a 
temperature in the cesium reservoir of 90$^\circ$C, at which the optical absorption length is greater than 80 $\mu$m. 
At this temperature reabsorption effects were still negligible, but the fluorescence signal was detectable even at low laser power.

\begin{figure*}[htpb]
	\includegraphics[width=\textwidth]{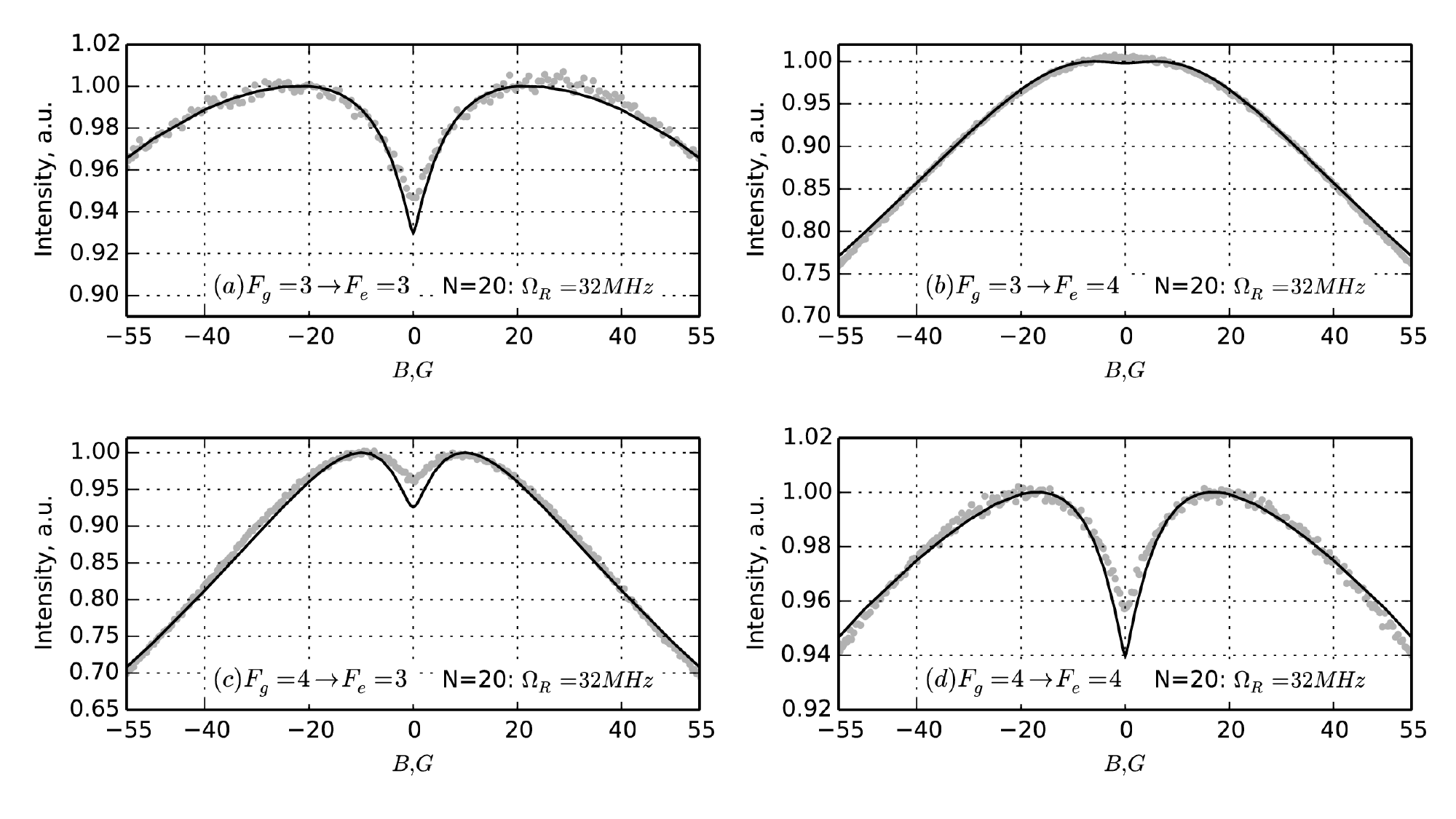}
	\caption{Magnetic resonances for different $D_1$ transitions at laser power of 160 $\mu W$ and wall separation $L=\lambda$. (grey dots - experiment, solid line - theory.}
	\label{fig:trans}
\end{figure*}

Figure \ref{fig:trans} shows the shapes of the fluorescence signal together with theoretical curves 
for all four $D_1$ transitions at a constant laser power of 160~$\mu$W and wall separation that is equal 
to the wavelength of the laser radiation. The laser beam diameter, laser power, laser wavelength, and cell temperature 
were directly measured from the experiment. The Rabi frequency was adjusted in order to obtain the best overall fit.  
The calculations were performed by dividing the beam into $N=20$ regions.

Analysing Fig. \ref{fig:trans}, as expected, we can see three dark resonances for transitions $F_g=3\rightarrow F_e =3$, and $F_g=4 \rightarrow F_e=3,4$. 
For the transition $F_g=3\rightarrow F_e=4$ a bright resonance would be expected in an ordinary cell, but the experimental and 
theoretical results show no bright resonance in the ETC. 
The bright resonance, for which the lowest contrast is expected, is destroyed by the rapid relaxation due to atom-wall collisions. 
This behaviour had been observed already experimentally in rubidium atoms confined to an ETC~\cite{Auzinsh:2010}.

\begin{figure*}[htpb]
	\includegraphics[width=\textwidth]{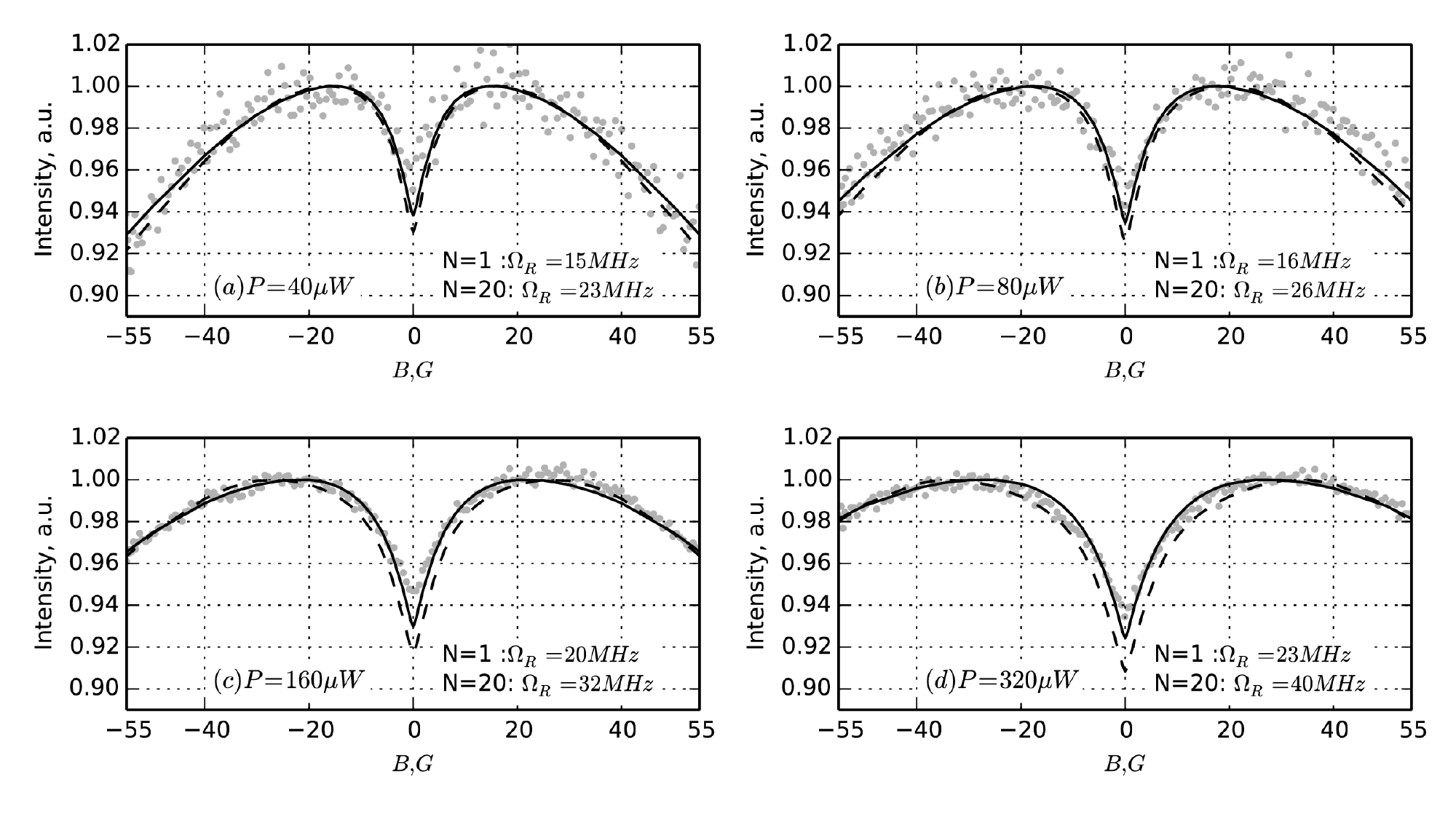}
	\caption{Power dependence of $F_g=3\rightarrow F_e=3$ transition compared to different theories (dots - experiment, dashed line - 1 region model, solid line - 20 region model).    }
	\label{fig:power}
\end{figure*}

Next we turned our attention to the power dependence of the resonance shapes. Figure \ref{fig:power} shows such a 
dependence for the $F_g=3\rightarrow F_e=3$ transition. To compare results of the improved description of transit relaxation with 
those obtained from a simpler treatment, 
two theoretical curves are given in each  subfigure. 
One curve has been calculated assuming that the laser beam was characterized by a single power density ($N=1$). 
To calculate the other curve, 
the laser beam  was assumed to be of Gaussian shape and split into twenty regions ($N=20$) of equal width (see Fig.~\ref{fig:dal1}). 
An extra averaging was performed over the values of the impact parameter of an atom crossing the beam off-axis. 
The number of regions was chosen in such a way that the results 
of theoretical calculations converged with further subdivisions resulting in negligible changes in the fluorescence signal. 
The same value of 32 MHz was used as the reduced Rabi frequency in the central region for all transitions.
The reduced Rabi frequency was modified for each transition according to its line strength. 
By comparing the results of the theoretical models, one can see that the theoretical curve of the 
single region model agrees well with experimental measurements
for cases were the laser power is low, but starts to deviate significantly in cases where 
the laser power is higher.
This behavior has already been observed experimentally numerous times in ordinary cells 
(\cite{Auzinsh:2008, Auzinsh:2012, Auzinsh:2013,Auzinsh_crossing:2013}) where theoretical calculations 
start to deviate from experimental data at 
laser powers where absorption at the center of the laser beam has reached saturation. 
In the case of the ETC even the lowest laser power used is noticeably higher than laser powers used in ordinary cell experiments, 
but, because of the strong relaxation by collisions with cell walls, noticeable deviations from experimental results start at only 
very high laser power densities. At such powers, by including the laser beam profile in the theoretical model, 
the agreement between theoretical calculations and experimental results could be greatly improved, especially in the description 
of the broader structure at larger magnetic field values (see~\cite{Auzinsh:2010}). 

\begin{figure*}[htp]
	\includegraphics[width=\textwidth]{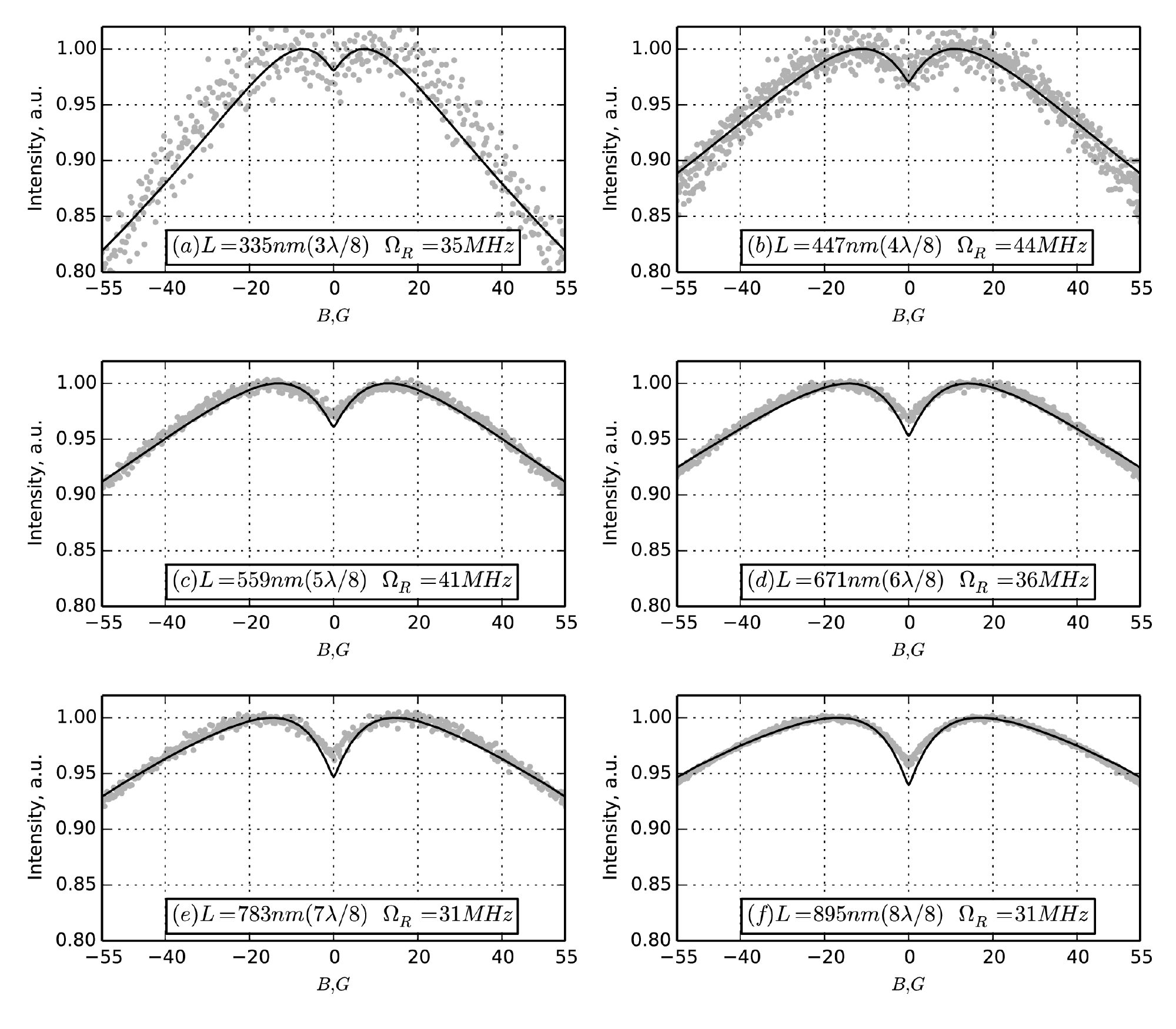}
	\caption{Magneto-optical resonance signal shape at different ETC cell thickness for transition $F_g=4\rightarrow F_e=4$.(dots - experiment, solid line - theory)}
	\label{fig:thickness} 
\end{figure*}

Finally we examined the signal dependence on cell thickness $L$. 
The experimental results for $F_g=4\rightarrow F_e=4$ together with theoretical calculations are shown in 
Fig.~\ref{fig:thickness}. The calculations were performed for $N=20$ concentric regions, 
and the Rabi frequencies for theoretical calculations where fitted for each cell thickness. 
The reason for tuning the Rabi frequencies was the fact that different thickness measurements 
where performed in different parts of the cell, for which the cell wall transparency may have been somewhat different. 

It can be seen that the  contrast of the zero-field resonance changes only by 
about 2\% while increasing the power density by a factor of 8 (Fig.~\ref{fig:power}). 
At the same time it can be seen that the contrast of the magneto-optical signal is governed by the temperature (Fig.~\ref{fig:temp_dep}) 
and cell thickness (Fig.~\ref{fig:thickness}) rather than by the laser power. 

\section{\label{Conclusions:level1}Conclusions}
ETCs are different from ordinary cells, because collisions with walls play a much larger role in contrast to ordinary cells, 
where transit relaxation effects usually dominate over wall-to-wall collisions even at moderate laser powers. 
In order to improve the 
theoretical description of magneto-optical signals in ETCs, the theoretical model was extended to include 
the intensity distribution in the cross-section of the laser beam
and to treat carefully collisions. As a result, measured nonlinear magneto-optical resonances 
for the Cs $D_1$ line for magnetic field values up to at least 55~G can be described much more precisely than before. 
Such a model could be very useful in improving the description of magneto-optical signals in ordinary cells at 
laser power densities that may be found in typical experiments.

\begin{acknowledgments}
The Riga group gratefully acknowledges financial support from the Latvian State Research Programme (VPP) project IMIS$^2$.
\end{acknowledgments}

\bibliography{rubidium}

\end{document}